%% file: KDD.tex
\documentclass[sigconf]{acmart}

\usepackage{booktabs} % For formal tables
\usepackage{booktabs}
\usepackage{algorithm}
\usepackage{algpseudocode}
\usepackage{amsmath}
\usepackage{amssymb}
\usepackage{array}
\usepackage{balance}
\usepackage[pangram]{blindtext}
\usepackage{caption}
\usepackage{color}

\usepackage{comment}
\usepackage{dblfloatfix}
\usepackage[inline]{enumitem}
\usepackage{epsfig}
\usepackage{etoolbox}
\usepackage{fancybox}
\usepackage{filecontents}
\usepackage{fixltx2e}
\usepackage{float}
\usepackage{graphicx}
\usepackage{threeparttable}
\usepackage{multirow}
\usepackage{multicol}
\definecolor{red}{rgb}{1,0,0}
\definecolor{green}{rgb}{0,1,0}
\definecolor{blue}{rgb}{0,0,1}

\setcopyright{none}

\makeatletter
\renewcommand\@formatdoi[1]{\ignorespaces}
\makeatother

\begin{document}
\title[\small{Mobile Security Enhancement against Adversarial Deception}]{DoPa: A Comprehensive CNN Detection Methodology \\against Physical Adversarial Attacks\\}
%\titlenote{Produces the permission block, and
%  copyright information}
%\subtitle{Extended Abstract}
%\subtitlenote{The full version of the author's guide is available as
%  \texttt{acmart.pdf} document}

\settopmatter{printacmref=false} % Removes citation information below abstract
\renewcommand\footnotetextcopyrightpermission[1]{} % removes footnote with conference information in first column
\pagestyle{plain} % removes running headers

%\author{Zirui XU}
%%\authornote{Dr.~Trovato insisted his name be first.}
%%\orcid{1234-5678-9012}
%\affiliation{%
%  \institution{George Mason University}
%  \streetaddress{4400 University Drive}
%  \city{Fairfax}
%  \state{Virginia}
%  \postcode{22030}
%}
%\email{zxu21@gmu.edu}
%
%\author{Fuxun Yu}
%%\authornote{The secretary disavows any knowledge of this author's %actions.}
%\affiliation{%
%  \institution{George Mason University}
%  \streetaddress{4400 University Drive}
%  \city{Fairfax}
%  \state{Virginia}
%  \postcode{22030}
%}
%\email{fyu2@gmu.edu}
%
%\author{Zirui Xu$\dagger$, Fuxun Yu$\dagger$, Chenchen Liu$\ddagger$, Xiang Chen$\dagger$}
%\affiliation{
%	$\dagger$George Mason University, Fairfax, Virginia, $\{$zxu21, fyu2, xchen26$\}$@gmu.edu
%	\\$\ddagger$Clarkson University, Potsdam, New York, chliu@clarkson.edu
%\vspace{1.5mm}
%}
\vspace{-3mm}
\author{Zirui Xu, Fuxun Yu, Xiang Chen}
\affiliation{
	George Mason University, Fairfax, Virginia\\ $\{$zxu21, fyu2, xchen26$\}$@gmu.edu
\vspace{1.5mm}
}

\renewcommand{\shortauthors}{Z. Xu \textit{et al.}}

%\begin{abstract}
%This paper provides a sample of a \LaTeX\ document which conforms,
%somewhat loosely, to the formatting guidelines for
%ACM SIG Proceedings.\footnote{This is an abstract footnote}
%\end{abstract}

%\begin{CCSXML}
%<ccs2012>
%  <concept_id>10010520.10010553.10010562</concept_id>
%  <concept_desc>Computer systems organization~Embedded systems</concept_desc>
%  <concept_significance>500</concept_significance>
%  <concept_id>10010520.10010575.10010755</concept_id>
%  <concept_desc>Computer systems organization~Redundancy</concept_desc>
%  <concept_significance>300</concept_significance>
%  <concept_id>10010520.10010553.10010554</concept_id>
%  <concept_desc>Computer systems organization~Robotics</concept_desc>
%  <concept_significance>100</concept_significance>
%  <concept_id>10003033.10003083.10003095</concept_id>
%  <concept_desc>Networks~Network reliability</concept_desc>
%  <concept_significance>100</concept_significance>
%</ccs2012>
%\end{CCSXML}

%\ccsdesc[500]{Computer systems organization~Embedded systems}
\vspace{-2mm}
\input{0_abstract}


\graphicspath{{}}
\maketitle

%\vspace{-1mm}
\input{1_introduction}
\input{2_preliminary}

\input{3_theory}
\input{4_image}
\input{5_audio}

\input{6_experiment}

\input{7_conclusion}

%\vspace{-2mm}
\bibliographystyle{ACM-Reference-Format}

\bibliography{KDD}

\end{document}

%% file: 0_abstract.tex
\vspace{-4mm}
\begin{abstract}

% Recently, Convolutional Neural Networks (CNNs) demonstrate considerable vulnerability to adversarial attacks, which inject particular perturbation into input data and mislead CNN prediction results.
Recently, Convolutional Neural Networks (CNNs) demonstrate a considerable vulnerability to adversarial attacks, which can be easily misled by adversarial perturbations.
	%Adversarial attacks usually inject in the the input data The traditional adversarial attacks mainly focus on algorithm domain which can only perturb digit input data.
With more aggressive methods proposed, adversarial attacks can be also applied to the physical world, causing practical issues to various CNN powered applications.
	To secure CNNs, adversarial attack detection is considered as the most critical approach. 
	% Therefore, it's critical to detect the adversarial attacks during CNN inference.
However, most existing works focus on superficial patterns and merely search a particular method to differentiate the adversarial inputs and natural inputs, ignoring the analysis of CNN inner vulnerability. Therefore, they can only target to specific physical adversarial attacks, lacking expected versatility to different attacks.
To address this issue, we propose \textit{DoPa} -- a comprehensive CNN detection methodology for various physical adversarial attacks. 
	% By interpreting the CNN vulnerability with activation visualization, we identify the vulnerability as its uncontrollability of the inconsistency between practical input patterns from physical adversarial attacks and the expected semantic patterns from prediction results.
	By interpreting the CNN's vulnerability, we find that non-semantic adversarial perturbations can activate CNN with significantly abnormal activations and even overwhelm other semantic input patterns' activations.
	Therefore, we add a self-verification stage to analyze the semantics of distinguished activation patterns, which improves the CNN recognition process.  
	% to detect and defense the physical adversarial attacks with only one CNN forward process involved.
	% We first identify two inconsistencies in input level and activation level respectively, which can be used to detect the physical adversarial attacks. 
	% % Then we propose the metrics to measure the two inconsistencies. 
	% When the non-semantic abnormal activation is detected, we can further recover the data by image painting and abnormal activation suppression.   
	% Based on the detection result, we further propose a data recovery methodology to defend the physical adversarial attacks. 
	% We apply the self-verification stage for both CNN based image and audio recognition applications.
	We apply such a detection methodology into both image and audio CNN recognition scenarios.
Experiments show that \textit{DoPa} can achieve an average rate of 90\% success for image attack detection and 92\% success for audio attack detection.
	% Moreover, the proposed defense methods are at most 2.3$\times$ faster compared to the state-of-the-art detection methods, making them feasible to resource-constrained platforms, such as mobile devices.
	% in image and audio respectively.

\hspace{-4mm}\textbf{Announcement:}[The original DoPa draft on arXiv was modified and submitted to a conference already, while this short abstract was submitted only for a presentation at the KDD 2019 AIoT Workshop.]
\end{abstract}

%% file: 1_introduction.tex
\vspace{-2mm}
\section{Introduction}
\label{sec:intr}

In the past few years, Convolutional Neural Networks (CNNs) have been widely applied in various cognitive applications, such as image classification~\cite{wang2017residual,zoph2018learning} and speech recognition~\cite{chorowski2015attention,chiu2018state}, \textit{etc}.
	Although effective and popular, CNN powered applications are facing a critical challenge -- adversarial attacks. 
	By injecting particular perturbations into input data, adversarial attacks can mislead CNN recognition results. 
	% The perturbations generated by traditional adversarial attacks are fragile, and can only be added into the digital data. 
	% Therefore, they can hardly threaten the recognition systems which obtain input data from the real world. 
	With aggressive methods proposed, adversarial perturbations can be concentrated into a small area and attached to the real objects, which easily threaten the CNN recognition systems in the physical world.
	% With aggressive methods proposed, adversarial perturbations can be concentrated into a small area and be attached to the real objects, which can be easily applied to the physical world.
	% which become more robust to external impacts such as environmental noise and camera angle shifting. 
	% Therefore, these adversarial attacks can be applied to the physical world. 
	% One of the most representative examples is the physical adversarial attack, which can bring real-world threats into these security-critical CNN applications. 
	% By injecting some well-crafted physical adversarial noise into original input, the CNN-based application will be deceived.
	 % and its attention will shift from original semantic natural input patterns to condensed non-semantic noise patterns.
% These kinds of physical adversarial attacks can threaten both image and audio CNN based applications. 
% Fig.~\ref{Threat_Model} shows a physical adversarial attack example on traffic signs.  
	% When we attach a well-crafted adversarial patch on the original stop sign, the traffic sign detection system will be misled to a wrong recognition result as a speed limit sign.
	% When user's mobile device capture an image or audio, the CNN-based application can extract the semantic natural input pattern on image/audio and get the correct recognition result as "bird" and "open". 
	% This misclassification reveals CNN's intrinsic vulnerability to physical adversarial attack. 
	Recently, physical adversarial attacks becomes severe with increasing CNN based applications~\cite{cauley2006nsa,boloor2019simple}.

	% discard original semantic pattern and cater to overwhelming activations with non-semantic patterns, which finally lead to a deception result ("airplane" and "close"). 
	% Recently, the scale and impact of such adversarial deception is significantly exacerbating with increasing CNN-based mobile applications and more practical adversarial deception methods involved~\cite{xx}.

Many research works have been proposed to detect the physical adversarial attacks~\cite{hayes2018visible,naseer2019local,yang2018characterizing,osadchy2017no}. 
% However, most of them merely focus on finding particular few features to differentiate the adversarial inputs and natural inputs, or they simply adopt multiple CNNs to conduct the cross-verification~\cite{multiversion,yang2018characterizing}.
	However, most of them neglect analysis of CNN's intrinsic vulnerability to adversarial attacks. 
	% Instead, some of them focus on eliminating explicit perturbation patterns from inputs which will introduce a considerable data processing cost~\cite{naseer2019local,xx}. 
	% Others simply adopt multiple CNNs to conduct the cross-verification which lack expected versatility to different physical adversarial attacks~\cite{multiversion,xx}. 
	Instead, either they merely focus on examining the superficial input patterns to differentiate the adversarial inputs and natural inputs or they simply adopt multiple CNNs to conduct the cross-verification~\cite{multiversion,yang2018characterizing}. 
	Therefore, all these methods have a certain drawback: they can only detect specific physical adversarial attacks, lacking versatility to different physical adversarial attacks.
In this paper, we propose \textit{DoPa}, a comprehensive detection methodology for physical adversarial attacks.  
	% By interpreting the CNN vulnerability with activation visualization methods~\cite{erhan2009visualizing, xu2015show}, we reveal that the CNN decision-making process lacks necessary qualitative semantics distinguishing ability. 
	By interpreting CNN's vulnerability, we reveal that the CNN's decision-making process lacks necessary qualitative semantics distinguishing ability: the non-semantic input patterns can dramatically activate CNN and overwhelm other semantic input patterns. 
	% it can be significantly activated by non-semantic input patterns and overwhelm other semantic input patterns. 
	% As a result, they cannot discriminate the inconsistency between practical input patterns from physical adversarial attack and the expected semantic patterns from prediction results.
	% As a result, they cannot discriminate the inconsistency between practical activation sources from physical adversarial attack and the expected semantic patterns from prediction results.
		% \textcolor{red}{As a result, they cannot discriminate the inconsistency between the input activation and corresponding semantics. }
	% To solve this problem, we improve CNN recognition process by adding a self-verification stage to verify whether the input contains adversarial perturbation.
	We improve the CNN recognition process by adding a self-verification stage to verify the semantics of distinguished activation patterns during CNN inference.  
	% to detect and defense the physical adversarial attacks with only one CNN forward process involved.
	% We first identify two inconsistencies in input level and activation level respectively, which can be used to detect the physical adversarial attacks. 
	% % Then we propose the metrics to measure the two inconsistencies. 
	% When the non-semantic abnormal activation is detected, we can further recover the data by image painting and abnormal activation suppression.   
	% Based on the detection result, we further propose the data recovery method to defense the physical adversarial attacks. 
	% By explicitly identifying the expected semantic patterns for each prediction class, we can verify whether they are inconsistent with practical activation source for each CNN forward process. 
	% Fig.~\ref{Threat_Model} illustrates the self-verification stage for a traffic sign adversarial attack. 
	% For each input image, the verification stage will locate the significant activation sources (shown in green circle) and calculate the input semantic inconsistency with the expected semantic patterns (shown in the right circle) according to the prediction result. 
		For each input image, the verification stage will locate the significant activation sources and calculate the input semantic inconsistency with the expected semantic patterns according to the prediction result. 
	% Once the inconsistency exceeds a pre-defined threshold, CNN will conduct a data recovery method to recover the input image. 
	Once the inconsistency exceeds a pre-defined threshold, CNN will predict the input is an adversarial example. 
	% Finally, we can get the correct recognition result. 
	% Our defense methodology depends on only one CNN inference with minimum computation components involved, which can be extended to both CNN based image and audio recognition applications.

	% After forward process, CNN can derive the practical input patterns by using neural network attention method and compare it with expected semantic patterns from predicted result. 
	% By introducing an inconsistency metric, CNN measure the inconsistency between its practical input patterns and expected semantic patterns. THerefore, it can identify whether the input is adversarial and can further eliminate the impact of adversarial attack, which can be well extended to both CNN based image and audio recognition applications.  

Specifically, we have the following contributions in this work:
% \vspace{-1.8mm}
\begin{itemize}
	\vspace{-1mm}
	\item By interpreting CNN's vulnerability, we discover that the non-semantic input patterns can significantly activate CNN and overwhelm other semantic input patterns. 
	% \vspace{-0.1mm}
	% \item We then identify the inconsistencies between the adversarial attack and normal recognition process with respect to the semantic and activation. We further propose the activation/semantic inconsistency metric to measure the them.
	\item We propose a self-verification stage to analyze and detect the abnormal activation patterns' semantics. Specifically, we develop the inconsistency between the local input patterns that cause the distinguished activations and the synthesized patterns with expected semantics. 

	\item We apply the proposed self-verification detection methodology into two scenarios for image and audio applications. 
	 % based on adversarial pattern removing and abnormal activation suppressing, respectively.  
		% \vspace{-1mm}
	% \item We design the experiments and evaluate the enhancement performance of our proposed method.
	% $bullet$ We propose a dynamic network reconfiguration scheme, which can selectively compute the network components to meet the dynamic resource constraints introduced by real-time mobile applications without retraining.
\end{itemize}
			\vspace{-1mm}
Experiments show that our method can achieve average 90\% and 92\% detection successful rates for image physical adversarial attacks and audio physical adversarial attacks, respectively.  
% Moreover, our methods are at most 2.3$\times$ faster than the state-of-the-art defense methods, which is feasible to various resource-constrained platforms, such as mobile devices. 

% \begin{figure}[t]
% 	\centering
% 	\captionsetup{justification=centering}
% 	\vspace{0mm}
% 	\includegraphics[width=3.1in]{Threat_Model}
% 	\vspace{-44mm}
% 	\caption{Physical Adversarial Attack for Traffic Sign}
% 	\vspace{-5mm}
% 	\label{Threat_Model}
% \end{figure}

%% file: 2_preliminary.tex
\vspace{-3mm}
\section{Background and Related Works}
\label{sec:prel}
\vspace{-1.5mm}

\subsection{Physical Adversarial Attacks}

% \begin{figure}[t]
% 	\centering
% 	\captionsetup{justification=centering}
% 	\vspace{-4.5mm}
% 	\includegraphics[width=3.0in]{Filterpruning}
% 	\vspace{-20mm}
% 	\caption{Structural Filter Pruning Overview}
% 	\vspace{-5mm}
% 	\label{Filterpruning}
% \end{figure}

The adversarial attack started to arouse researchers' general concern with adversarial examples, which were first introduced by~\cite{goodfellow2014explaining}.
	The adversarial examples were designed to project prediction errors into input space to generate noises, which can perturb digital input data (\textit{e.g.}, images and audio clips) and manipulate prediction results. 
	Since then, various adversarial attacks were proposed, such as L-BFGS~\cite{szegedy2013intriguing}, FGSM~\cite{goodfellow2014explaining}, CW~\cite{carlini2017towards}, \textit{etc}. 
	Most of them share a similar mechanism, which tries to cause the most error increment within model activation and regulate the noises within the input space.

Recently, such an attack approach was also brought from the algorithm domain into the physical world, which were referred as physical adversarial attacks. 
	~\cite{eykholt2017robust} first leveraged a masking method to concentrate adversarial perturbations into a small area and implement attacks on real traffic signs with taped graffiti. 
	~\cite{brown2017adversarial} then extended the scope of physical attacks with adversarial patches. 
	With more aggressive image patterns than taped graffiti, these patches could be attached to physical objects arbitrarily and have a certain degree of model transferability. 

Beyond aforementioned image scenarios, some physical adversarial attacks also have been proposed to audios. 
	Yakura \textit{et al.}~\cite{yakura2018robust} proposed an audio physical adversarial attack that can still be effective after playback and recording in the physical world. 
	~\cite{yakura2018robust} generated audio adversarial command in a normal song which can be played through the air.  

	Compared to the noise based adversarial attacks, these physical adversarial attacks reduce the attack difficulty and further impair the practicality and reliability of deep learning technologies.

% \begin{figure}[t]
% 	\centering
% 	\captionsetup{justification=centering}
% 	\vspace{-0mm}
% 	\includegraphics[width=3.3in]{Semantic_Mismatch}
% 	\vspace{-33mm}
% 	\caption{Visualized Neuron's Input Pattern by Activation Maximization Visualization}
% 	\vspace{-6mm}
% 	\label{Semantic_Mismatch}
% \end{figure}

\vspace{-4mm}
\subsection{Image and Audio physical Adversarial Attack Detection}

% \begin{figure}[t]
% 	\centering
% 	\captionsetup{justification=centering}
% 	\vspace{-4.5mm}
% 	\includegraphics[width=3.0in]{Image_Adv}
% 	\vspace{-11mm}
% 	\caption{Typical Image Recognition Process and Physical Adversarial Deception}
% 	\vspace{-3mm}
% 	\label{Image_Adv}
% \end{figure}

% \textcolor{red}{Fig.~\ref{Image_Adv} illustrate a typical image recognition process and its physical adversarial deception. 
% By attaching the physical adversarial patches on the original input image, the CNN will be mislead to a wrong recognition result. }

There are several works have been proposed to detect such physical adversarial attacks in the image and speech recognition process~\cite{hayes2018visible,multiversion,yang2018characterizing,naseer2019local}. 
	% Naseer \textit{et al.} proposed a local gradients smoothing scheme against physical adversarial attacks~\cite{naseer2019local}. By regularizing gradients in the estimated noisy region before feeding images into CNN for inference, their method can eliminate the potential impacts from adversarial attacks. 
	Hayes \textit{et al.} proposed a physical image adversarial attack detection method by applying image erosion and dilation~\cite{hayes2018visible}. With these image processing methods, they successfully detect the localization of adversarial noises in input images.
	Zeng \textit{et al.} leveraged multiple Automatic Speech Recognition (ASR) systems to detect audio physical adversarial attack according to the cross-verification methodology~\cite{multiversion}.
	% However, their method lacks certain versatility which cannot detect the adversarial attacks with model transferability.  
	Yang \textit{et al.} proposed an audio adversarial attack detection method by exploring the temporal dependency existing in audio adversarial attacks~\cite{yang2018characterizing}.
	% However, their method requires multiple CNN recognition inferences which is time-consuming. 

% Although these methods are effective for physical adversarial attacks defection, they still have certain disadvantages regarding versatility.  
% For example, local gradients smoothing requires the manipulation for each pixel of the input image, which will introduce a large number of computation workload. 
Although having effective detection performance for physical adversarial attacks, these methods are designed for solving specific adversarial attack which are not integrated for different physical adversarial attack situations.
Therefore, we develop a comprehensive defection methodology to address the above issue.

%% file: 3_theory.tex
\vspace{-3.5mm}
\section{CNN Vulnerability Analysis for Physical Adversarial Attacks}
\label{sec:algo}

% In the introduction, Fig.~\ref{Threat_Model} has demonstrated opposite characteristics in terms of size, semantics, and activation strength between adversarial attack input and original input. 
% This opposition implies certain CNN design flaws in the decision-making process.

% In this section, we first interpret CNN's vulnerability with activation maximization visualization method~\cite{erhan2009visualizing}.   
% By leveraging this vulnerability, the physical adversarial attacks can significantly activate the deception-class related neurons and overwhelm the correct neurons, which can mislead the result.
% Based on the vulnerability interpretation, we identify the inconsistency between practical activation sources from physical adversarial attacks and the expected semantic patterns from prediction results.
% \textcolor{red}{Based on the vulnerability interpretation, we identify the inconsistency between input activation and corresponding semantics.}
% Based on the vulnerability analysis, we identify two inconsistencies that can be used to detect the physical adversarial attacks. 
% We propose a metric that can measure such inconsistencies and further fix CNN's vulnerability by adding a self-verification stage in the inference.

In this section, we first interpret the CNN vulnerability by analyzing input patterns' semantics with the activation maximization visualization~\cite{erhan2009visualizing}. 
Based on the semantics analysis, we identify adversarial attack patches as non-semantic input patterns with abnormally distinguished activations. 
Specifically, to evaluate semantics, we propose metrics that can measure the inconsistencies between the local input patterns that cause the distinguished activations and the synthesized patterns with expected semantics.
% Specifically, to evaluate the semantics, we propose two metrics: input semantic inconsistency and prediction activation inconsistency that can measure the inconsistencies in input patterns and prediction activations between 
Based on the inconsistency analysis, we further propose a detection methodology by adding a self-verification stage into CNN inference. 
% methodology to detect and defense the physical adversarial attacks.  

% \begin{figure}[t]
% 	\centering
% 	\captionsetup{justification=centering}
% 	\vspace{-0mm}
% 	\includegraphics[width=3.3in]{Semantic_Mismatch}
% 	\vspace{-33mm}
% 	\caption{Visualized Neuron's Input Pattern by Activation Maximization Visualization}
% 	\vspace{-6mm}
% 	\label{Semantic_Mismatch}
% \end{figure}

\begin{figure}[t]
	\centering
	\captionsetup{justification=centering}
	\vspace{0mm}
	\includegraphics[width=3.3in]{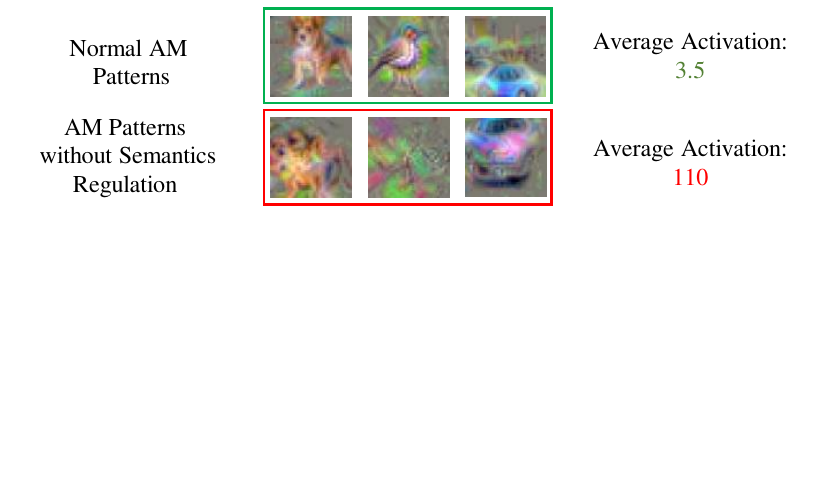}
	\vspace{-33mm}
	\caption{Visualized Neuron's Input Pattern by Activation Maximization Visualization}
	\vspace{-6mm}
	\label{Semantic_Mismatch}
\end{figure}

\vspace{-3.5mm}
% \subsection{CNN Vulnerability Interpretation with \textbf{Activation/Semantic Inconsistency}}
\subsection{CNN Vulnerability Interpretation}

\hspace{3.5mm}\textbf{\textit{Interpretation and Assumption:}}
In a typical image or audio recognition process, CNN extracts features from the original input data and gradually derive a prediction result.
However, when injecting physical adversarial perturbations into the original data, CNN will be misled to a wrong prediction result. 
% To better interpret the vulnerability, we first analyze CNN's vulnerability by using a typical image physical adversarial attack -- adversarial patch attack as an example.
% Through Fig.~\ref{Threat_Model}, we find an phenomenon: CNN's recognition process will be easily misled when adding some concentrated adversarial patterns on the original input data. 
To better interpret the CNN vulnerability, we first use a typical image physical adversarial attack -- adversarial patch attack as an example.
% Comparing to original input data which has specific semantics, physical adversarial perturbation such as adversarial patch demonstrates no semantic. 
In Fig.~\ref{Threat_Model}, by comparing with the original input, we find an adversarial patch usually has no constraints in color/shape, \textit{etc}. 
Such patches usually sacrifice the semantic structures so as to cause significant abnormal activations and overwhelm other input patterns' activations. 
% These condensed physical adversarial perturbation  
% Adversarial perturbation and original image demonstrate opposite characteristic in terms of size and semantics. 
% This opposition in semantics implies certain CNN vulnerability in the decision-making process. 
% Since typical physical adversarial perturbations such as adversarial patches are condensed noise patches without semantics. 
\textit{Therefore, we make an \textbf{assumption} that CNN lacks qualitative semantics distinguishing ability which can be activated by the non-semantic adversarial patch during inference.}

\textbf{\textit{Assumption Verification:}}
To verify such an assumption, we need to investigate the semantic of each neuron in CNN. 
According to our assumption, the non-semantic input patterns will lead to abnormal activations while the semantic input patterns generate normal activations.
We adopt a visualized CNN semantic analysis method -- Activation Maximization Visualization (AM)~\cite{erhan2009visualizing}. 
	By generating a pattern $V(N_i^l)$ to maximize the $ith$ neuron $N_i^l$ in the layer of $l$, we can visualize each neuron's most activated input.
Mathematically, this process can be formulated as:
\vspace{-2.5mm}
\small
\begin{equation}
	\medmuskip=-1mu
	V(N_i^l)=\mathop{\arg\max}_{X} A_i^l(X), \qquad X \leftarrow X + \eta \cdot \frac{\partial A_i^l(X)}{\partial X},
	\label{eq:am}
	\vspace{-2.5mm}
\end{equation}
\normalsize
where, $A_i^l(X)$ is the activation of $N_i^l$ from an input image X, $\eta$ is the gradient ascent step size.
\begin{comment}
	AM can generate a pattern to visualize each neuron's most activated semantic input. 
}
}
The generation process of pattern $V(N_i^l)$ can be considered as synthesizing an input image to a CNN model that delicately maximizes the activation of the $ith$ neuron $N_i^l$ in the layer of $l$. Specifically, this process can be formulated as:
\vspace{-1.5mm}
\small
\begin{equation}
	\medmuskip=-1mu
	V(N_i^l)=\mathop{\arg\max}_{X} A_i^l(X), \qquad X \leftarrow X + \eta \cdot \frac{\partial A_i^l(X)}{\partial X}
	\label{eq:am}
	\vspace{-1.5mm}
\end{equation}
\normalsize
where, $A_i^l(X)$ is the activation of $N_i^l$ from an input image X, $\eta$ is the gradient ascent step size.
\end{comment}

Fig.~\ref{Semantic_Mismatch} shows the visualized neurons' semantic input patterns by using AM. 
As the traditional AM method is designed for semantics interpretation, many feature regulations and hand-engineered natural image references are involved in generating interpretable visualization patterns. 
% The clear objects indicate th
Therefore we can get three AM patterns with an average activation magnitude value of 3.5 in Fig.~\ref{Semantic_Mismatch} (a). 
The objects in the three patterns indicate they have clear semantics. 
% We can find that each of these three patterns has clear object. 
However, when we remove these semantics regulations in the AM process, we obtain three different visualized patterns as shown in Fig.~\ref{Semantic_Mismatch} (b).
% which shows \textbf{inconsistencies} with previous ones. 
We can find that these three patterns are non-semantic, but they have significant abnormal activations with an average magnitude value of 110.
% but they can achieve much larger activation magnitude for three neurons. 
This phenomenon can prove our assumption that CNN neurons lack semantics distinguishing ability and can be significantly activated by \textbf{non-semantic} inputs patterns. 
% Therefore, CNN decision-making process is mainly based on indiscriminate quantitative activations, lacking necessary qualitative semantics distinguishing ability. 

% \begin{figure}[b]
% 	\centering
% 	\captionsetup{justification=centering}
% 	\vspace*{-6mm}
% 	\includegraphics[width=2.8in]{Adversarial_Patch}
% 	\vspace{-3mm}
% 	\caption{Image Adversarial Patch Attack}
% 	\vspace{-1mm}
% 	\label{Adversarial_Patch}
% \end{figure}

% \begin{figure}[t]
% 	\centering
% 	\captionsetup{justification=centering}
% 	\vspace*{-3mm}
% 	\includegraphics[width=3.0in]{Adversarial_Patch}
% 	\vspace{-3mm}
% 	\caption{Image Adversarial Patch Attack}
% 	\vspace{-8mm}
% 	\label{Adversarial_Patch}
% \end{figure}

\vspace{-2mm}
\subsection{\hspace{-1mm}Inconsistency Metrics for Input Semantic and Prediction Activation}

% We already analyzed the CNN's vulnerability to physical adversarial attacks. 
	% To identify whether CNN is under attacking, we aim to find the difference between the process of natural image recognition and physical adversarial attacks.

\hspace{3.5mm}\textbf{\textit{Inconsistency Identification:}}
% To identify the non-semantic input patterns for the attack detection, it is critical to identify the \textbf{inconsistencies} between the physical adversarial attacks and natural image recognition.
To identify the non-semantic input patterns for the attack detection, we examine its impacts during CNN inference by comparing the natural image recognition with the physical adversarial attacks.

% To identify the non-semantic input patterns for the attack detection, we aim to compare the natural image recognition with the physical adversarial attacks.
	% Next, we are aiming to find some features that can identify 

Fig.~\ref{Adversarial_Patch} shows a typical adversarial patch based physical attack.
The patterns in the left circles are the primary activation sources from the input images, and the bars on the right are the neurons' activations in the last convolutional layer.
% The impacts of adversarial patches can be demonstrated as \textbf{inconsistencies} at two levels:
% % From the aspect of \textbf{input semantic patterns}, 
% From the aspect of input patterns, we identify \textbf{input semantic consistency} between the adversarial patch and primary activation source on the original image;
% From the aspect of prediction activations, we identify \textbf{prediction activation consistency} between the adversarial input and the original input, which are their activation magnitudes.
% The impacts of adversarial patches can be demonstrated as \textbf{inconsistencies} at two levels:
% From the aspect of \textbf{input semantic patterns}, 
From the aspect of input patterns, we find a significant difference between the adversarial patch and primary activation source on the original image, which is referred as \textbf{\textit{input semantic inconsistency}};
From the aspect of prediction activation magnitudes, we observe another difference between the adversarial input and the original input, which is referred as \textbf{\textit{prediction activation inconsistency}}.
% Therefore, we formulate the inconsistencies at two levels:
% Therefore, the impacts of adversarial patches can be demonstrated as \textbf{inconsistencies} at two levels:
% We identify two inconsistencies in terms of input level and activation level that can help us to detect the adversarial patches. 	
% We can find that: 
% (1) With natural image, CNN model can successfully learn to pay attention to semantic-meaningful locations with the main object and activate the expected class-related neurons with reasonable degrees. (2) However, with the adversarial patch injected, the deception-class related neurons are significantly activated and quickly overwhelm the correct neurons, and therefore mislead the prediction results. 
% We find two inconsistencies in terms of input level and activation level which can help us to detect the adversarial patches.
% We refer these two inconsistencies as \textbf{Semantic/Activation Inconsistency}.
% We refer the two inconsistencies as \textbf{Semantic/Activation Inconsistency}.  

\textbf{\textit{Inconsistency Metrics:}}\
We further define two metrics to indicate above two inconsistencies' degrees. 

{\textbf{\textit{1)Input Semantic Inconsistency Metric.}}
This metric measures the input semantic inconsistency between the non-semantic adversarial patches and the semantic local input patterns from the natural image. 
It can be defined as follows: 
\vspace{-1mm}
\small
\begin{equation}
	\medmuskip=-1mu
	\thinmuskip=-1mu
	\thickmuskip=-1mu
	% D(I_{pra},I_{exp})=1-S(I_{pra},I_{exp}), I_{pra} \xleftarrow{\Re} \sum_{X,Y}{A(x,y)}, I_{exp} \xleftarrow{\Re} \sum_{X,Y}{A'(x,y)},
	D(P_{pra},P_{ori})=1-S(P_{pra},P_{ori}), P_{pra} \xleftarrow{\Re} \Phi:{A_i^l(p)}, P_{ori} \xleftarrow{\Re} \Phi:{A_i^l(o)},
	\label{eq:metric1}
	% \vspace{-0.5mm}
\end{equation}
\normalsize
where $P_{pra}$ and $P_{ori}$ represent the input patterns from the adversarial input and the original input. 
$\Phi:{A_i^l(p)}$ and $\Phi: {A_i^l(o)}$ represent the set of neurons' activations produced by adversarial patch and the original input, respectively. $\Re$ maps neurons' activations to the primary local input patterns.  
$S$ represents a similarity metric. 
% such as Pearson Similarity~\cite{benesty2009pearson}. 

\textit{\textbf{2)Prediction Activation Inconsistency Metric.}}
The second inconsistency is in the activation level, which reveals the activations' magnitude distribution inconsistency in the last convolutional layer between the adversarial input and the original input. 
% Compared with the activation magnitude distribution of original natural image, the adversarial patch will cause abnormal activation magnitude distribution.
We also use a similar metric to measure it as follows: 
\vspace{-1mm}
\small
\begin{equation}
	\medmuskip=-1mu
	\thinmuskip=-1mu
	\thickmuskip=-1mu
	% D(I_{pra},I_{exp})=1-S(I_{pra},I_{exp}), I_{pra} \xleftarrow{\Re} \sum_{X,Y}{A(x,y)}, I_{exp} \xleftarrow{\Re} \sum_{X,Y}{A'(x,y)},
	D(f_{pra},f_{ori})=1-S(f_{pra},f_{ori}), f_{pra} \sim \Phi:{A_i^l(p)}, f_{ori} \sim \Phi:A_i^l(o),
	% f_{pra} \xleftarrow{\Re} \Phi:{A_i^l(P)}, f_{exp} \xleftarrow{\Re} \Phi:{A_i^l(O)},
	\label{eq:metric1}
	% \vspace{-0.5mm}
\end{equation}
\normalsize
where $f_{pra}$ and $I_{ori}$ represent the magnitude distribution of activations in the last convolutional layer generated by the adversarial input and the original input data. 

	% Different from images, the audio data requires more processing efforts. 
	% As Fig.~\ref{Audio_Adv} shows, during the audio recognition, the input waveform needs to pass Mel-frequency Cepstral Coefficient (MFCC) conversion to be transferred from the time domain into the time-frequency domain. 
	% % Therefore, by applying CAM, we can only obtain the input pattern in the MFCC feature level with less semantics.
	% Therefore, the original input audio data will loss semantics after the MFCC conversion. 
	% Therefore, we only leverage the activation inconsistency to detect the audio physical adversarial attacks.  
	% We can find that: 
	% Therefore, we extend the the Practical Input/Expected Semantics Inconsistency can be obtained from activation distribution in the last layer we mentioned above.

% $\Re$ maps neurons' activations to the activation magnitude distribution.  $\Phi:{A_i^l(P)}$ and $\Phi: {A_i^l(O)}$ represent the set of neurons' activations produced by practical input and the original input, respectively.  
% $S$ represents a similarity metric and $\Re$ is a mapping relationship.

For the above two inconsistency metrics, we can easily obtain $P_{pra}$ and $f_{pra}$ since they come from the input data. 
However, $P_{ori}$ and $f_{ori}$ are not easily to get because of the variety of the natural input data. 
Therefore, we need to synthesize the standard input data which can provide the semantic input patterns and activation magnitude distribution.   
The synthesized input data for each prediction class can be obtained from a standard dataset. 
By feeding CNN with a certain number of input from the standard dataset, we can record the average activation magnitude distribution in last convolutional layer. 
Moreover, we can locate the primary semantic input patterns for each prediction class. 

\vspace{-3mm}
\subsection{Inconsistency for Physical Adversarial Attack Detection}

% With the two activation semantics inconsistency metrics we built above, we can  
% The above metric reflects the inconsistency between the practical input and expected prediction input. 
% According to our analysis, the inconsistence calculated from our metric for an adversarial attack input will be larger than a natural input.    
% The larger inconsistency value indicate 
% In this part, we propose our physical adversarial attack defense method based on the above analysis and metric. 
% Based on the above analysis, we propose our physical adversarial attack defense methodology, which can fix CNN's vulnerability by adding a self-verification stage in the decision-making process. 
% With this self-verification stage, CNN can detect and eliminate the physical adversarial attacks by itself in one shot with only one CNN inference involved. 

The proposed two inconsistencies demonstrate the difference between physical adversarial attacks and natural image recognition in terms of input patterns and prediction activations, we can leverage the two inconsistency metrics to measure the inconsistencies and further detect the physical adversarial attacks. Specifically, when the measured inconsistency values exceed the given thresholds, we consider the input as an adversarial input. 

Since the two metrics can be easily calculated during CNN inference, we integrate them as a \textbf{\textit{self-verification}} stage in the CNN decision-making process. During the inference, CNN leverages this self-verification stage to measure the inconsistencies and further achieve adversarial input verification.   

% Based on the above analysis, we propose a detection methodology which applying a self-verification stage in the CNN decision-making process:
% By comparing the semantic/activation inconsistency between practical input and ground truth, this self-verification stage can detect and defense the physical adversarial attacks in one-shot. 
% (1) We first feed the input into the CNN inference and obtain the prediction class.
% (2) Next, CNN can locate the primary activation sources from the practical input and obtain the activations in the last convolutional layer.  
% (3) Then CNN leverages the input semantic inconsistency metric and the prediction activation inconsistency metric to measure the two inconsistencies between the practical input and the synthesized data with the prediction class.  
% % \textcolor{red}{The ground-truth of the prediction class include two elements.  First one is the semantic input patterns while the second one is   }
% (4) Once inconsistency exceeds the given threshold, CNN will consider the input as an adversarial input.

We apply such \textbf{\textit{detection}} methodology for both image and audio applications in Section 4 and Section 5.

\begin{figure}[t]
	\centering
	\captionsetup{justification=centering}
	\vspace*{-5mm}
	\includegraphics[width=3.0in]{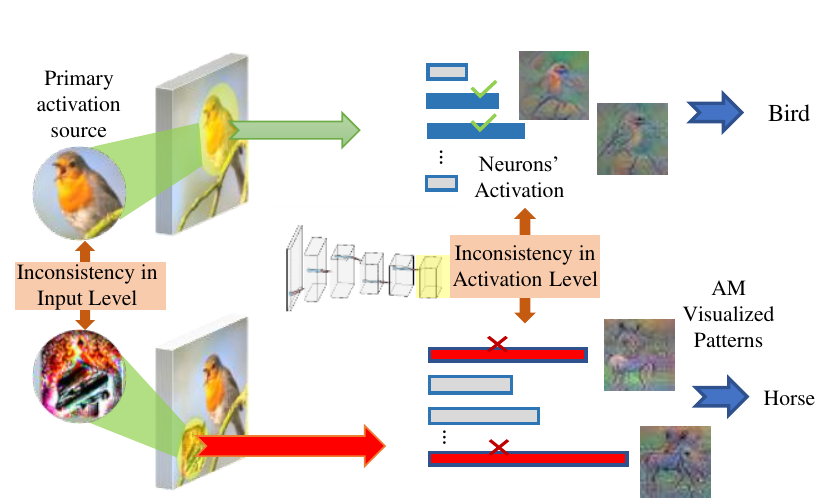}
	\vspace{-3mm}
	\caption{Image Adversarial Patch Attack}
	\vspace{-8mm}
	\label{Adversarial_Patch}
\end{figure}

%% file: 4_image.tex
\vspace{-2mm}
\section{Image Physical Adversarial Attack Detection}
\label{sec:imag}

% In the last section, we propose the defense methodology which consists of the self-verification stage and data recovery.
In this section, we will describe how to leverage our proposed methodology to detect the image physical adversarial attacks based on the \textbf{\textit{input semantic inconsistency}} in input pattern level. 
% proposed detection methodology for image physical adversarial attacks. 
% As physical attacks leverage CNNs' inherent flaws for semantic reasoning, CNN robustness enhancement against physical attacks might be achieved at the cost of significant model structure redesign. Thus, to guarantee defense effectiveness and generality, we design our defense method based on the scheme of ``detection-as-defense".
% For image physical adversarial attacks defense, we mainly depend on the \textbf{input semantic inconsistency} in input pattern level. 

\textbf{\textit{Inconsistency Derivation:}}
First, by using Class Activation Mapping (CAM)~\cite{zhou2016learning}, we locate the primary activation source from the input image. 
Assume the $k^{th}$ activation in the last convolutional layer is $A_k(x,y)$ and its spatial location is $(x,y)$, we can compute a weight sum of all activations at the $(x,y)$ in the last convolutional layer as: 
	\vspace{-2mm} 
\small
\begin{equation}
	\medmuskip=-1mu
	A_{T}(x,y)=\sum^{1}_{K}A_k(x,y),
	\label{eq:cam}
	\vspace{-1.5mm}
\end{equation}
\normalsize
where $K$ is the total number of activations in the last convolutional layer. 
A larger value of $A_{T}(x,y)$ represents the activation source in the input image at the corresponding spatial location $(x,y)$ is more important for classification result.
A specific input semantic inconsistency metric is further required for achieving CNN self-verification. 
Since the input adversarial patch contains more high-frequency information than the natural semantic input patterns, we differentiate them from the aspect of frequency. 2D Fast Fourier Transform (2D-FFT)~\cite{ayres2008measuring} is introduced to concentrate the low-frequency component together in frequency domain.
Furthermore, we convert the frequency-domain pattern to a binary pattern with an adaptive threshold. 
For binary patterns, we can observe the significant difference between adversarial input and semantic synthesized input. 
Therefore, we replace $S(I_{pra},I_{ori})$ as Jaccard Similarity Coefficient (JSC)~\cite{niwattanakul2013using} and propose our image adversarial attack inconsistency metric as: 
\small
\begin{equation}
	\medmuskip=-2mu
	\thinmuskip=-1.5mu
	\thickmuskip=-2mu
	D(P_{pra},P_{exp})=1-JSC(P_{pra},P_{exp})=\frac{|P_{pra}\bigcup P_{exp}|-|P_{pra}\bigcap P_{exp}|}{|P_{pra}\bigcup P_{exp}|},
	\label{eq:metric1}
	\vspace{-1mm}
\end{equation}
\normalsize
where $I_{exp}$ is the synthesized semantic pattern with predicted class.  
$P_{pra}\bigcap P_{exp}$ means the numbers of pixels where the pixel value of $P_{pra}$ and $P_{exp}$ both equal to 1.

\textbf{\textit{Self-Verification:}}
% With the derived inconsistency metric, we propose our specific detection methodology which contains the self-verification stage.
With the derived inconsistency metric, we propose our specific detection methodology which contains the self-verification stage.
% With the above inconsistency metric, we apply our proposed defense methodology which contains 6 steps from image inputting to image recovery.
The entire process of our method is described in Fig.~\ref{Image_Patch}. 
% \textbf{Self-verification for Detection}
For each input image, we apply CAM to locate the source location of the biggest model activations. Then we crop the image to obtain the patterns with maximum activations.
In the step of semantic test, we calculate the consistency between $I_{pra}$ and $I_{exp}$. Once the inconsistency value is higher than a predefined threshold, we consider an adversarial input detected.
\begin{comment}
\textbf{Data Recovery for Image}
After the patch is detected, we do the image data recovery by directly removing patch from the original input data. To eliminate the attack effects, we further leverage image inpainting technology to repair the image such as image interpolation~\cite{bertalmio2000image}.  
At last, we feed back the recovery image into CNN to do the prediction again. 
\end{comment}
% (1) The testing image is first fed into CNN. 
% (2) We apply CAM to detect the source location of the biggest model activation from the input image. 
% (3) Based on CAM result, we can crop the image to obtain the patterns with maximum activation. 
% (4) In this step, we will do the semantic test between maximum activation pattern and ground-truth semantic pattern. Concretely, we take the maximum activation pattern in local input as $I_{pra}$. Then we calculate the consistency between $I_{pra}$ and $I_{exp}$. Once the inconsistency is higher than a predefined threshold, we consider a significant inconsistency detected. 
% We could assume the captured input pattern is an adversarial attack input. 
% (5) After the patch is detected, we can directly remove it from the original input data and repair the image using image inpainting~\cite{bertalmio2000image}. 
% (6) At last, we feed back the recovery image into CNN to do the prediction again. 
% With the above 6 steps, we can detect and further defense an image physical adversarial attack input during CNN inference process. We will evaluate its performance in our experiment. 
With above steps, we can detect the image physical adversarial attack through self-verification stage during CNN inference process.

%% file: 5_audio.tex
\vspace{-2mm}
\section{Audio Physical Adversarial Attack Detection}
\label{sec:audi}

In this section, we will introduce the detailed detection flow for audio physical adversarial attacks. 

\textbf{\textit{Inconsistency Derivation:}}
We leverage the \textbf{prediction activation inconsistency} to detect audio physical adversarial attacks since the original input audio will loss semantics after the Mel-frequency Cepstral Coefficient (MFCC) conversion. 
\begin{comment}
	Different from images, the audio data requires more processing efforts. 
	During the audio recognition, the input waveform needs to pass Mel-frequency Cepstral Coefficient (MFCC) conversion to be transferred from the time domain into the time-frequency domain. 
	% Therefore, by applying CAM, we can only obtain the input pattern in the MFCC feature level with less semantics.
	In that case, the original input audio data will loss semantics after the MFCC conversion. 
	Therefore, we leverage the \textbf{prediction activation inconsistency} to detect audio physical adversarial attacks.  
\end{comment}
	% We can find that: 
	% Therefore, we extend the the Practical Input/Expected Semantics Inconsistency can be obtained from activation distribution in the last layer we mentioned above.
% As aforementioned, we focus on the activation inconsistency in the audio aspect. 
More specifically, we measure the activation magnitude distribution inconsistency between the practical input and the synthesized data with the same prediction class. 
We adopt a popular similarity evaluation method - Pearson Correlation Coefficient (PCC)~\cite{benesty2009pearson} and the inconsistency metric is defined as: 
\small
\begin{equation}
	\medmuskip=-3mu
	\thinmuskip=-3mu
	\thickmuskip=-3mu
	D(f_{pre},f_{exp})=1-PCC(f_{pre},f_{exp})=1-\frac{E[(f_{pre}-\mu_{pre})(f_{exp}-\mu_{exp})]}{\sigma_{pra}\sigma_{exp}},
	\label{eq:metric2}
	\vspace{-1.5mm}
\end{equation}
\normalsize
where $I_{pre}$ and $I_{exp}$ represent the activations in the last convolutional layer for both practical input and synthesized input.  
$\mu_a$ and $\mu_o$ denote the mean values of $f_{pre}$ and $f_{exp}$, $\sigma_{pra}$ and $\sigma_{exp}$ are the standard derivations, and $E$ means the overall expectation. 

\textbf{\textit{Self-Verification:}}
With derived inconsistency metric, we further apply self-verification stage to CNN for the audio physical adversarial attack. 
% (1) The starting MFCC feature is difficult for divergence measurement represented, since it is a high-dimensional time-frequency matrix. However, after the CNN process, the MFCC feature will abstracted and represented as standard activation patterns. These activation patterns provide us a good chance to conduct divergence measurement. 
% (2) Due to noise generation constraint, the adversarial noises are designed to manipulate the final prediction layer’s activation, while the original activation patterns in the hidden layers are kept to a greater extent. Therefore, we could utilize the discrepancy between the hidden-layers and predication layers to reflect the deception and detect the adversarial noises.
% Same with image physical adversarial attack, the defense in audio is also based on `detection-as-defense" methodology. 
The detection flow is described as following: 
	We first obtain activations in the last convolutional layer for every possible input word by testing CNN with a standard dataset. 
	Then we calculate the inconsistency value $D(I_{pra},I_{exp})$.
	 % between the activated hidden layer pattern and the ground-truth activation pattern corresponding to the prediction result. 
	If the model is attacked by the audio adversarial attack, $D(I_{pra},I_{exp})$ will exceed a pre-defined threshold. 
	% According to our preliminary experiments tested with various attacks, there exists a large range for the threshold to distinguish the natural and the adversarial audio, which can benefit our accurate detection.
	According to our preliminary experiments which are tested with various attacks, $D(I_{pra},I_{exp})$ of an adversarial input is usually larger than 0.18 while a natural input's $D(I_{pra},I_{exp})$ is usually smaller than 0.1. Therefore, there exists a large range for the threshold to distinguish the natural and the adversarial input audios, which can benefit our accurate detection. 

	% there exists a large range for the threshold to distinguish the natural and the adversarial audio. $D(I_{pra},I_{exp})$ of an adversarial input is usually larger than 0.18 while a natural input's $D(I_{pra},I_{exp})$ is usually smaller than 0.1. This 
	% as we will show in experiment.

\begin{figure}[t]
	\centering
	\captionsetup{justification=centering}
	\vspace{0mm}
	\includegraphics[width=3.3in]{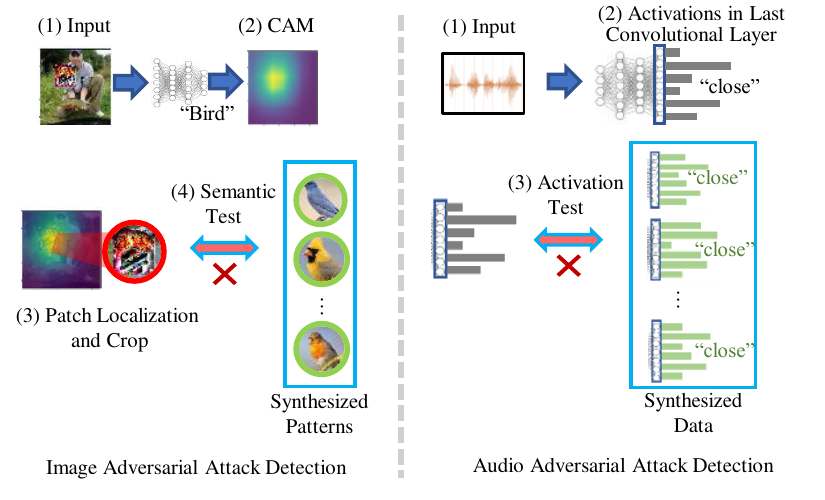}
	\vspace{-3mm}
	\caption{Detection Process for Image and Audio Physical Adversarial Attacks }
	\vspace{-8mm}
	\label{Image_Patch}
\end{figure}

%% file: 6_experiment.tex
\vspace{-2mm}
\section{Experiment and Evaluation}
\label{sec:Expe}

In this section, we evaluate our method's effectiveness in terms of image and audio physical adversarial attacks. 
	% The CNN models and datasets used in our experiments are listed in Table~\ref{tab:1}: 
	For physical adversarial attack in image scenarios, we test our detection method's performance on Inception-V3~\cite{Szeg:2015:CVPR}, VGG-16~\cite{Simo:2014:arXiv}, and ResNet-18~\cite{He:2016:CVPR} using ImageNet dataset~\cite{deng2009imagenet}.
	For audio scenarios, we use Command Classification Model~\cite{morgan2001speech} on Google Voice Command dataset~\cite{morgan2001speech}. 
	%Other three state-of-the-art methods are used as a comparison~\cite{multiversion,yang2018characterizing,rajaratnam2018noise}.

\vspace{-2mm}
\subsection{CNN Image Physical Adversarial\\ Attack Detection Evaluation}

% In this part, we evaluate our proposed detection method for the image physical adversarial attack scenario.  
Our detection method is mainly evaluated for adversarial patch attacks. 
	The adversarial patches are generated by using Inception-V3 as the base model. The generated patches with high transferability are utilized to attack three models: Inception-V3 itself and two other models, namely VGG-16 and ResNet-18.
	Then we apply our detection method on the attacks for three models and test their detection success rates. The baseline methods is \textit{Blind}, which is one state-of-the-art detection method~\cite{hayes2018visible}. And the threshold for inconsistency is set as 0.46.  
Table~\ref{tab:2} shows the overall detection performance. \textit{Blind} achieves an average rate of 87\% success for detection while \textit{DoPa} demonstrate an average 90\% detection successful rate.  
	On all three models, \textit{DoPa} consistently shows higher detection success rate than~\cite{hayes2018visible}. 
	% The further proposed image recovery could help to correct predictions, resulting in 80.3\%$\sim$82\% accuracy recovery improvement on different models while \textit{Blind} only achieves 78.2\%$\sim$79.5\% accuracy recovery improvement. 
	% In terms of efficiency, the process time cost of our detect method for one physical adversarial attack is from 67\textit{ms}$\sim$71\textit{ms} while the \textit{Blind} is from 132\textit{ms}$\sim$153\textit{ms}. 
By the above comparison, we show that our detection method has better detection performance than \textit{Blind} \textit{w.r.t} detection effectiveness.

% \begin{figure}[t!]
% 	\centering
% 	\captionsetup{justification=centering}
% 	\vspace{-14mm}
% 	\includegraphics[width=3.3in]{Audio_Defense}
% 	\vspace{-3mm}
% 	\caption{Audio Adversarial Attack Defense }
% 	\vspace{-6mm}
% 	\label{Audio_Defense}
% \end{figure}

\vspace{-2mm}
\subsection{CNN Audio Physical Adversarial\\ Attack Detection Evaluation}

% In this part, we evaluate the effectiveness of the proposed detection method in audio physical adversarial attack scenarios.
% According to our preliminary experiments tested with various attacks, there exists a large range for the threshold to distinguish the natural and the adversarial audio, which can benefit our accurate detection.
Since there exists a large range for the inconsistency threshold to distinguish the natural and the adversarial audio, we leverage the grid search to find the best one and set as 0.11 in this experiment. 
	% The inconsistency threshold for adversarial detection is obtained by the grid search and set as 0.11 in this experiment.
For comparison, we re-implement another two state-of-the-art detection methods: \textit{Dependency Detection}~\cite{yang2018characterizing} and \textit{Multiversion}~\cite{multiversion}. Four methods~\cite{goodfellow2014explaining,kurakin2016adversarial,carlini2017towards,alzantot2018did} are used as attacking methods to prove the generality of our detection method.
Table~\ref{tab:3} shows the overall performance comparison results. 

\textit{DoPa} can always achieve more than 92\% detection success rate for all types of audio physical adversarial attacks. By contrast, \textit{Dependency Detection} achieves 89\% detection success rate in average while \textit{Multiversion Detection} only has average 74\%.
Therefore, \textit{DoPa} demonstrates a best detection accuracy.

\small
\begin{table}[t]
	\vspace{0mm}
	\centering
	\caption{Image Adversarial Attack Detection Evaluation}
	\vspace{-4mm}
\setlength{\tabcolsep}{0.7mm}{
\begin{tabular}{|c|c|c|c|c|c|c|}
	\hline
	\multirow{2}{*}{Model/Method} & \multicolumn{2}{c|}{Inception-V3} & \multicolumn{2}{c|}{VGG-16} & \multicolumn{2}{c|}{ResNet-18} \\ \cline{2-7} 
                                & Blind*       & Ours                & Blind*    & Ours             & Blind*      & Ours              \\ \hline
Detection Succ. Rate            & 88\%          & \textbf{91\%}       & 89\%       & \textbf{90\%}    & 85\%         & \textbf{89\%}     \\ \hline
\end{tabular}}
	\begin{tablenotes}
		\scriptsize
		\item[1] *:Blind~\cite{hayes2018visible}
	\end{tablenotes}
	\label{tab:2}
	\vspace{-4mm}
\end{table}
\normalsize

\small
\begin{table}[t]
	\vspace{0mm}
	\centering
	\caption{Audio Adversarial Attack Detection Evaluation}
	\vspace{-4mm}
\setlength{\tabcolsep}{0.7mm}{
\begin{tabular}{|c|c|c|c|c|}
\hline
Method                 & FGSM  & BIM & CW  & Genetic \\ \hline
Dependency Detection   & 91\% & 89\% & 90\% & 88\%     \\ \hline
Multiversion Detection & 81\%   & 73\% & 79\% & 68\%     \\ \hline
Ours                   & 95\%   & 94\% & 94\% & 92\%     \\ \hline
\end{tabular}}
	\label{tab:3}
	\vspace{-4mm}
\end{table}
\normalsize

\begin{comment}
\small
\begin{table}[t]
	\vspace{1mm}
	\centering
	\caption{Audio Adversarial Attack Data Recovery Evaluation}
	\vspace{-4mm}
\setlength{\tabcolsep}{0.5mm}{
\begin{tabular}{|c|c|c|c|c|c|}
\hline
Method                                                         & FGSM~\cite{goodfellow2014explaining}         & BIM~\cite{kurakin2016adversarial}          & CW~\cite{carlini2017towards}          & Genetic~\cite{alzantot2018did}      & Time Cost      \\ \hline
No Recovery                                                     & 10\%          & 5\%           & 4\%           & 13\%          & NA             \\ \hline
\begin{tabular}[c]{@{}c@{}}Dependency\\ Detection~\cite{yang2018characterizing} \end{tabular} & 85\%          & 83\%          & 80\%          & 80\%          & 1813ms         \\ \hline
Noise Flooding~\cite{rajaratnam2018noise}                                                 & 62\%          & 65\%          & 62\%          & 59\%          & 1246ms         \\ \hline
Ours                                                           & \textbf{87\%} & \textbf{88\%} & \textbf{85\%} & \textbf{83\%} & \textbf{521ms} \\ \hline
\end{tabular}}
% \scriptsize
% \raggedright $*$Original \textit{LeNet} Computation Cost Baseline: Capacity:42.77M   Memory: 12.6MB   Energy:3.56mJ\\
% \vspace{-0.5mm}
% \raggedright $*$Original \textit{VGG-13} Computation Cost Baseline: Capacity:317.04M   Memory: 62MB   Energy:33.06mJ
% \normalsize
	\label{tab:3}
	\vspace{-6mm}
\end{table}
\normalsize

\end{comment}

%% file: 7_conclusion.tex
	\vspace{-2mm}
\section{Conclusion}
\label{sec:conc}

In this paper, we propose a CNN detection methodology for physical adversarial attacks for both image and audio recognition applications. 
Leveraging the comprehensive CNN vulnerability analysis and two novel CNN inconsistency metrics, our method can effectively detect both image and audio physical adversarial attacks with average 90\% and 92\% detection successful rates.

%% file: KDD.bbl

\begin{thebibliography}{29}

%%% the Web address.

\ifx \showCODEN    \undefined \def \showCODEN     #1{\unskip}     \fi
\ifx \showDOI      \undefined \def \showDOI       #1{#1}\fi
\ifx \showISBNx    \undefined \def \showISBNx     #1{\unskip}     \fi
\ifx \showISBNxiii \undefined \def \showISBNxiii  #1{\unskip}     \fi
\ifx \showISSN     \undefined \def \showISSN      #1{\unskip}     \fi
\ifx \showLCCN     \undefined \def \showLCCN      #1{\unskip}     \fi
\ifx \shownote     \undefined \def \shownote      #1{#1}          \fi
\ifx \showarticletitle \undefined \def \showarticletitle #1{#1}   \fi
\ifx \showURL      \undefined \def \showURL       {\relax}        \fi
% invisible to TeX
\providecommand\bibfield[2]{#2}
\providecommand\bibinfo[2]{#2}
\providecommand\natexlab[1]{#1}
\providecommand\showeprint[2][]{arXiv:#2}

\bibitem[\protect\citeauthoryear{Alzantot and \emph{et al.}}{Alzantot and
  \emph{et al.}}{2018}]%
        {alzantot2018did}
\bibfield{author}{\bibinfo{person}{M. Alzantot} {and} \bibinfo{person}{\emph{et
  al.}}} \bibinfo{year}{2018}\natexlab{}.
\newblock \showarticletitle{Did You Hear that? Adversarial Examples Against
  Automatic Speech Recognition}.
\newblock \bibinfo{journal}{\emph{arXiv preprint arXiv:1801.00554}}
  (\bibinfo{year}{2018}).
\newblock


\bibitem[\protect\citeauthoryear{Ayres and \emph{et al.}}{Ayres and \emph{et
  al.}}{2008}]%
        {ayres2008measuring}
\bibfield{author}{\bibinfo{person}{C. Ayres} {and} \bibinfo{person}{\emph{et
  al.}}} \bibinfo{year}{2008}\natexlab{}.
\newblock \showarticletitle{Measuring Fiber Alignment in Electrospun Scaffolds:
  A User's Guide to the 2D Fast Fourier Transform Approach}.
\newblock \bibinfo{journal}{\emph{Journal of Biomaterials Science, Polymer
  Edition}} \bibinfo{volume}{19}, \bibinfo{number}{5} (\bibinfo{year}{2008}),
  \bibinfo{pages}{603--621}.
\newblock


\bibitem[\protect\citeauthoryear{B and \emph{et al.}}{B and \emph{et
  al.}}{2019}]%
        {boloor2019simple}
\bibfield{author}{\bibinfo{person}{Adith B} {and} \bibinfo{person}{\emph{et
  al.}}} \bibinfo{year}{2019}\natexlab{}.
\newblock \showarticletitle{Simple physical adversarial examples against
  end-to-end autonomous driving models}.
\newblock \bibinfo{journal}{\emph{arXiv preprint arXiv:1903.05157}}
  (\bibinfo{year}{2019}).
\newblock


\bibitem[\protect\citeauthoryear{Benesty and \emph{et al.}}{Benesty and
  \emph{et al.}}{2009}]%
        {benesty2009pearson}
\bibfield{author}{\bibinfo{person}{J. Benesty} {and} \bibinfo{person}{\emph{et
  al.}}} \bibinfo{year}{2009}\natexlab{}.
\newblock \showarticletitle{Pearson Correlation Coefficient}.
\newblock In \bibinfo{booktitle}{\emph{Noise reduction in speech processing}}.
  \bibinfo{publisher}{Springer}, \bibinfo{pages}{1--4}.
\newblock


\bibitem[\protect\citeauthoryear{Brown and \emph{et al.}}{Brown and \emph{et
  al.}}{2017}]%
        {brown2017adversarial}
\bibfield{author}{\bibinfo{person}{T. Brown} {and} \bibinfo{person}{\emph{et
  al.}}} \bibinfo{year}{2017}\natexlab{}.
\newblock \showarticletitle{Adversarial Patch}.
\newblock \bibinfo{journal}{\emph{arXiv preprint arXiv:1712.09665}}
  (\bibinfo{year}{2017}).
\newblock


\bibitem[\protect\citeauthoryear{Carlini and \emph{et al.}}{Carlini and
  \emph{et al.}}{2017}]%
        {carlini2017towards}
\bibfield{author}{\bibinfo{person}{N. Carlini} {and} \bibinfo{person}{\emph{et
  al.}}} \bibinfo{year}{2017}\natexlab{}.
\newblock \showarticletitle{Towards Evaluating the Robustness of Neural
  Networks}. In \bibinfo{booktitle}{\emph{\textit{Proc. of SP}}}.
  \bibinfo{pages}{39--57}.
\newblock


\bibitem[\protect\citeauthoryear{Chiu and \emph{et al.}}{Chiu and \emph{et
  al.}}{2018}]%
        {chiu2018state}
\bibfield{author}{\bibinfo{person}{C. Chiu} {and} \bibinfo{person}{\emph{et
  al.}}} \bibinfo{year}{2018}\natexlab{}.
\newblock \showarticletitle{State-of-the-art Speech Recognition with
  Sequence-to-sequence Models}. In \bibinfo{booktitle}{\emph{\textit{Proc. of
  ICASSP}}}. \bibinfo{pages}{4774--4778}.
\newblock


\bibitem[\protect\citeauthoryear{Chorowski and \emph{et al.}}{Chorowski and
  \emph{et al.}}{2015}]%
        {chorowski2015attention}
\bibfield{author}{\bibinfo{person}{J. Chorowski} {and}
  \bibinfo{person}{\emph{et al.}}} \bibinfo{year}{2015}\natexlab{}.
\newblock \showarticletitle{Attention-based Models for Speech Recognition}. In
  \bibinfo{booktitle}{\emph{\textit{Proc. of NIPS}}}.
  \bibinfo{pages}{577--585}.
\newblock


\bibitem[\protect\citeauthoryear{Deng and \emph{et al.}}{Deng and \emph{et
  al.}}{2009}]%
        {deng2009imagenet}
\bibfield{author}{\bibinfo{person}{J. Deng} {and} \bibinfo{person}{\emph{et
  al.}}} \bibinfo{year}{2009}\natexlab{}.
\newblock \showarticletitle{Imagenet: A large-scale Hierarchical Image
  Database}. In \bibinfo{booktitle}{\emph{\textit{Proc. of CVPR}}}.
  \bibinfo{pages}{248--255}.
\newblock


\bibitem[\protect\citeauthoryear{Erhan and \emph{et al.}}{Erhan and \emph{et
  al.}}{2009}]%
        {erhan2009visualizing}
\bibfield{author}{\bibinfo{person}{D. Erhan} {and} \bibinfo{person}{\emph{et
  al.}}} \bibinfo{year}{2009}\natexlab{}.
\newblock \showarticletitle{Visualizing Higher-layer Features of A Deep
  Network}.
\newblock \bibinfo{journal}{\emph{University of Montreal}}
  \bibinfo{volume}{1341}, \bibinfo{number}{3} (\bibinfo{year}{2009}),
  \bibinfo{pages}{1}.
\newblock


\bibitem[\protect\citeauthoryear{Eykholt and \emph{et al.}}{Eykholt and
  \emph{et al.}}{2017}]%
        {eykholt2017robust}
\bibfield{author}{\bibinfo{person}{K. Eykholt} {and} \bibinfo{person}{\emph{et
  al.}}} \bibinfo{year}{2017}\natexlab{}.
\newblock \showarticletitle{Robust Physical-world Attacks on Deep Learning
  Models}.
\newblock \bibinfo{journal}{\emph{arXiv preprint arXiv:1707.08945}}
  (\bibinfo{year}{2017}).
\newblock


\bibitem[\protect\citeauthoryear{Goodfellow and \emph{et al.}}{Goodfellow and
  \emph{et al.}}{2014}]%
        {goodfellow2014explaining}
\bibfield{author}{\bibinfo{person}{I. Goodfellow} {and}
  \bibinfo{person}{\emph{et al.}}} \bibinfo{year}{2014}\natexlab{}.
\newblock \showarticletitle{Explaining and Harnessing Adversarial Examples}.
\newblock \bibinfo{journal}{\emph{arXiv preprint arXiv:1412.6572}}
  (\bibinfo{year}{2014}).
\newblock


\bibitem[\protect\citeauthoryear{Hayes}{Hayes}{2018}]%
        {hayes2018visible}
\bibfield{author}{\bibinfo{person}{J. Hayes}.} \bibinfo{year}{2018}\natexlab{}.
\newblock \showarticletitle{On Visible Adversarial Perturbations \& Digital
  Watermarking}. In \bibinfo{booktitle}{\emph{\textit{Proc. of CVPR}
  Workshops}}. \bibinfo{pages}{1597--1604}.
\newblock


\bibitem[\protect\citeauthoryear{He and \emph{et al.}}{He and \emph{et
  al.}}{2015}]%
        {He:2016:CVPR}
\bibfield{author}{\bibinfo{person}{K. He} {and} \bibinfo{person}{\emph{et
  al.}}} \bibinfo{year}{2015}\natexlab{}.
\newblock \showarticletitle{{Deep Residual Learning for Image Recognition}}. In
  \bibinfo{booktitle}{\emph{\textit{Proc. of CVPR}}}.
  \bibinfo{pages}{770--778}.
\newblock


\bibitem[\protect\citeauthoryear{James}{James}{2018}]%
        {cauley2006nsa}
\bibfield{author}{\bibinfo{person}{V. James}.} \bibinfo{year}{2018}\natexlab{}.
\newblock \bibinfo{title}{Google is making it easier than ever to give any app
  the power of object recognition}.
\newblock
\newblock
\urldef\tempurl%
\url{https://www.theverge.com/2017/6/15/15807096/google-mobile-ai-mobilenets-neural-networks}
\showURL{%
\tempurl}


\bibitem[\protect\citeauthoryear{Kurakin and \emph{et al.}}{Kurakin and
  \emph{et al.}}{2016}]%
        {kurakin2016adversarial}
\bibfield{author}{\bibinfo{person}{A. Kurakin} {and} \bibinfo{person}{\emph{et
  al.}}} \bibinfo{year}{2016}\natexlab{}.
\newblock \showarticletitle{Adversarial Examples in the Physical World}.
\newblock \bibinfo{journal}{\emph{arXiv preprint arXiv:1607.02533}}
  (\bibinfo{year}{2016}).
\newblock


\bibitem[\protect\citeauthoryear{Morgan and \emph{et al.}}{Morgan and \emph{et
  al.}}{2001}]%
        {morgan2001speech}
\bibfield{author}{\bibinfo{person}{S. Morgan} {and} \bibinfo{person}{\emph{et
  al.}}} \bibinfo{year}{2001}\natexlab{}.
\newblock \bibinfo{title}{Speech Command Input Recognition System for
  Interactive Computer Display with Term Weighting Means Used in Interpreting
  Potential Commands from Relevant Speech Terms}.
\newblock
\newblock
\newblock
\shownote{US Patent 6,192,343.}


\bibitem[\protect\citeauthoryear{Naseer and \emph{et al.}}{Naseer and \emph{et
  al.}}{2019}]%
        {naseer2019local}
\bibfield{author}{\bibinfo{person}{M. Naseer} {and} \bibinfo{person}{\emph{et
  al.}}} \bibinfo{year}{2019}\natexlab{}.
\newblock \showarticletitle{Local Gradients Smoothing: Defense against
  localized adversarial attacks}. In \bibinfo{booktitle}{\emph{\textit{Proc. of
  WACV}}}. \bibinfo{pages}{1300--1307}.
\newblock


\bibitem[\protect\citeauthoryear{Niwattanakul and \emph{et al.}}{Niwattanakul
  and \emph{et al.}}{2013}]%
        {niwattanakul2013using}
\bibfield{author}{\bibinfo{person}{S. Niwattanakul} {and}
  \bibinfo{person}{\emph{et al.}}} \bibinfo{year}{2013}\natexlab{}.
\newblock \showarticletitle{Using of Jaccard Coefficient for Keywords
  similarity}. In \bibinfo{booktitle}{\emph{\textit{Proc. of IMECS}}},
  Vol.~\bibinfo{volume}{1}. \bibinfo{pages}{380--384}.
\newblock


\bibitem[\protect\citeauthoryear{Osadchy and \emph{et al.}}{Osadchy and
  \emph{et al.}}{2017}]%
        {osadchy2017no}
\bibfield{author}{\bibinfo{person}{M. Osadchy} {and} \bibinfo{person}{\emph{et
  al.}}} \bibinfo{year}{2017}\natexlab{}.
\newblock \showarticletitle{No Bot Expects the DeepCAPTCHA! Introducing
  Immutable Adversarial Examples, With Applications to CAPTCHA Generation}.
\newblock \bibinfo{journal}{\emph{IEEE Transactions on Information Forensics
  and Security}} \bibinfo{volume}{12}, \bibinfo{number}{11}
  (\bibinfo{year}{2017}), \bibinfo{pages}{2640--2653}.
\newblock


\bibitem[\protect\citeauthoryear{Simonyan and \emph{et al.}}{Simonyan and
  \emph{et al.}}{2014}]%
        {Simo:2014:arXiv}
\bibfield{author}{\bibinfo{person}{K. Simonyan} {and} \bibinfo{person}{\emph{et
  al.}}} \bibinfo{year}{2014}\natexlab{}.
\newblock \showarticletitle{{Very Deep Convolutional Networks for Large-scale
  Image Recognition}}.
\newblock \bibinfo{journal}{\emph{arXiv preprint arXiv:1409.1556}}
  (\bibinfo{year}{2014}).
\newblock


\bibitem[\protect\citeauthoryear{Szegedy and \emph{et al.}}{Szegedy and
  \emph{et al.}}{2013}]%
        {szegedy2013intriguing}
\bibfield{author}{\bibinfo{person}{C. Szegedy} {and} \bibinfo{person}{\emph{et
  al.}}} \bibinfo{year}{2013}\natexlab{}.
\newblock \showarticletitle{Intriguing Properties of Neural Networks}.
\newblock \bibinfo{journal}{\emph{arXiv preprint arXiv:1312.6199}}
  (\bibinfo{year}{2013}).
\newblock


\bibitem[\protect\citeauthoryear{Szegedy and \emph{et al.}}{Szegedy and
  \emph{et al.}}{2015}]%
        {Szeg:2015:CVPR}
\bibfield{author}{\bibinfo{person}{C. Szegedy} {and} \bibinfo{person}{\emph{et
  al.}}} \bibinfo{year}{2015}\natexlab{}.
\newblock \showarticletitle{{Going Deeper with Convolutions}}. In
  \bibinfo{booktitle}{\emph{\textit{Proc. of CVPR}}}. \bibinfo{pages}{1--9}.
\newblock


\bibitem[\protect\citeauthoryear{Wang and \emph{et al.}}{Wang and \emph{et
  al.}}{2017}]%
        {wang2017residual}
\bibfield{author}{\bibinfo{person}{F. Wang} {and} \bibinfo{person}{\emph{et
  al.}}} \bibinfo{year}{2017}\natexlab{}.
\newblock \showarticletitle{Residual Attention Network for Image
  Classification}. In \bibinfo{booktitle}{\emph{\textit{Proc. of CVPR}}}.
  \bibinfo{pages}{3156--3164}.
\newblock


\bibitem[\protect\citeauthoryear{Yakura and \emph{et al.}}{Yakura and \emph{et
  al.}}{2018}]%
        {yakura2018robust}
\bibfield{author}{\bibinfo{person}{H. Yakura} {and} \bibinfo{person}{\emph{et
  al.}}} \bibinfo{year}{2018}\natexlab{}.
\newblock \showarticletitle{Robust Audio Adversarial Example for A Physical
  Attack}.
\newblock \bibinfo{journal}{\emph{arXiv preprint arXiv:1810.11793}}
  (\bibinfo{year}{2018}).
\newblock


\bibitem[\protect\citeauthoryear{Yang and \emph{et al.}}{Yang and \emph{et
  al.}}{2018}]%
        {yang2018characterizing}
\bibfield{author}{\bibinfo{person}{Z. Yang} {and} \bibinfo{person}{\emph{et
  al.}}} \bibinfo{year}{2018}\natexlab{}.
\newblock \showarticletitle{Characterizing Audio Adversarial Examples Using
  Temporal Dependency}.
\newblock \bibinfo{journal}{\emph{arXiv preprint arXiv:1809.10875}}
  (\bibinfo{year}{2018}).
\newblock


\bibitem[\protect\citeauthoryear{Zeng and \emph{et al.}}{Zeng and \emph{et
  al.}}{2018}]%
        {multiversion}
\bibfield{author}{\bibinfo{person}{Q. Zeng} {and} \bibinfo{person}{\emph{et
  al.}}} \bibinfo{year}{2018}\natexlab{}.
\newblock \showarticletitle{A Multiversion Programming Inspired Approach to
  Detecting Audio Adversarial Examples}.
\newblock \bibinfo{journal}{\emph{arXiv preprint arXiv:1812.10199}}
  (\bibinfo{year}{2018}).
\newblock


\bibitem[\protect\citeauthoryear{Zhou and \emph{et al.}}{Zhou and \emph{et
  al.}}{2016}]%
        {zhou2016learning}
\bibfield{author}{\bibinfo{person}{B. Zhou} {and} \bibinfo{person}{\emph{et
  al.}}} \bibinfo{year}{2016}\natexlab{}.
\newblock \showarticletitle{Learning Deep Features for Discriminative
  Localization}. In \bibinfo{booktitle}{\emph{\textit{Proc. of CVPR}}}.
  \bibinfo{pages}{2921--2929}.
\newblock


\bibitem[\protect\citeauthoryear{Zoph and \emph{et al.}}{Zoph and \emph{et
  al.}}{2018}]%
        {zoph2018learning}
\bibfield{author}{\bibinfo{person}{B. Zoph} {and} \bibinfo{person}{\emph{et
  al.}}} \bibinfo{year}{2018}\natexlab{}.
\newblock \showarticletitle{Learning Transferable Architectures for Scalable
  Image Recognition}. In \bibinfo{booktitle}{\emph{\textit{Proc. of CVPR}}}.
  \bibinfo{pages}{8697--8710}.
\newblock


\end{thebibliography}
